# Recoil Implantation Using Gas-Phase Precursor Molecules

Angus Gale[1], Johannes E. Fröch[1,*], Mehran Kianinia[1], James Bishop[1], Igor Aharonovich[1,2], Milos Toth[1,2,*]

[1]School of Mathematical and Physical Sciences, Faculty of Science, University of Technology Sydney, Ultimo, New South Wales 2007, Australia

[2]ARC Centre of Excellence for Transformative Meta-Optical Systems (TMOS), University of Technology Sydney, Ultimo, New South Wales 2007, Australia

Corresponding Author: johannes.froech@uts.edu.au, milos.toth@uts.edu.au

**ABSTRACT**

Ion implantation underpins a vast range of devices and technologies that require precise control over the physical, chemical, electronic, magnetic and optical properties of materials. A variant termed "recoil implantation" – in which a precursor is deposited onto a substrate as a thin film and implanted *via* momentum transfer from incident energetic ions – has a number of compelling advantages, particularly when performed using an inert ion nano-beam [Fröch *et al.,* Nat Commun 11, 5039 (2020)]. However, a major drawback of this approach is that the implant species are limited to the constituents of solid thin films. Here we overcome this limitation by demonstrating recoil implantation using gas-phase precursors. Specifically, we fabricate nitrogen-vacancy (NV) color centers in diamond using an $Ar^+$ ion beam and the nitrogen-containing precursor gases $N_2$, $NH_3$ and $NF_3$. Our work expands the applicability of recoil implantation to most of the periodic table, and to applications in which thin film deposition/removal is impractical.

# INTRODUCTION

Ion implantation techniques are commonly used to alter material properties for a wide range of applications. For instance, it has critically underpinned the advancement of semiconductor industries towards our technology standard today,[1] and research in fields such as solid state chemical engineering, optoelectronics, and quantum science have benefitted from the development of reliable implantation technologies.[2, 3] Due to their importance, there remains intense interest in improving implantation methods, particularly advances in mask-free, direct-write implantation at precisely-located sites, and fabrication of dopant gradients. For example, in the field of nano and quantum photonics, precise placement of optical dopants in photonic and optoelectronic devices is required, and deterministic ion implantation techniques are therefore highly sought after.

While conventional ion implanters offer a wide range of possible implantation energies and source ions, they are unable to achieve localized implantation on submicron scales without the need for masking procedures.[4, 5] Therefore, several methods are commonly used in material science, based on equipment beyond standard broad-beam ion implanters. This equipment includes focused ion beam (FIB) systems, which produce highly-focused, nanoscale beams. However, the ion species are limited by the ion source, which is typically a liquid metal ion source[6-8], a plasma source[9], or a gas field ionization source.[10] A recent demonstration has expanded this to a Paul trap, where ions are captured in an electrostatic trap and then accelerated towards a target,[11, 12] and techniques such as implantation through a pierced AFM tip have been used to achieve precise spatial localization.[13, 14] In addition, a more specialized technique was demonstrated recently using a scanning electron microscope and gas-phase precursor molecules.[15]

Further to the above, we have recently demonstrated the use of recoil implantation[16-18] in combination with a standard FIB system.[19] In that work, momentum transfer from inert ions in a nano-scale beam to a thin film was used for implantation of the film constituents. This was demonstrated by implanting group IV elements into a bulk diamond substrate, which resulted in the creation of group IV color centers. These centers have characteristic photoluminescence emissions,[20] which were distinctly identified by confocal photoluminescence measurements. The recoil implantation technique provides control over dopant density using the ion beam irradiation and scanning parameters, and enables ultra-shallow implantation, with the majority of dopants located within the first 2 nm of the surface.[19] Hence, this technique allows for a wide range of implant species using a single ion source, as well as beam-directed control over the location and density of the dopants. However, it is limited to the use of solid-state precursors that can be deposited in the form of a removable film. Here we eliminate this limitation by replacing solid thin films with gases that are injected into the system during FIB irradiation. We showcase this approach by creating NV centers[21] in a diamond substrate by using $N_2$, $NH_3$ and $NF_3$ as the implantation precursor species.

We note that whilst undesired recoil implantation of gas-phase oxygen impurities by broad ion beams has been observed previously[22], this process has not been used to engineer functional material properties such as the generation of NV spin defects in diamond demonstrated in the present work. Our results expand the technique of FIB-directed recoil implantation to encompass the vast majority of the periodic table, demonstrates the use of both inert and reactive precursor molecules, and negates the need for thin film deposition and removal steps that may be difficult depending on the properties of the substrate.

**RESULTS and DISCUSSION**

The method uses a standard dual beam microscope (DBM - Thermo Fisher Scientific Helios G4), a tool that is commonly available in microscopy and material science laboratories for purposes of cross sectioning, lamella preparation for transmission electron microscopy, and for nanofabrication.[23, 24] The same system is typically equipped with a gas injection system (GIS), where a capillary is placed within 500 μm of the substrate and used to locally deliver precursor molecules for gas-assisted nanofabrication, e.g. chemical vapor deposition or chemically-enhanced etching of the surface. Here, the same system is utilized for sub-surface implantation of gas molecule constituents, achieved by injecting and delivering nitrogen-based precursor gases to a local area of a substrate during FIB irradiation, as is schematically shown in Figure 1(a). As the gas molecules are injected and directed towards the sample, a proportion will adsorb on the surface. These gaseous molecules will either diffuse on the surface or undergo thermal or FIB-induced desorption, characterized by the diffusion path length and the mean desorption time, respectively.[25, 26] Within this timeframe, high energy primary ions can interact with the adsorbates, transfer momentum to the molecules, and thus implant their constituents into the substrate.

The substrate used in this work is electronic-grade diamond (N < 5 ppb) purchased from Element Six, cleaned prior to experiments by ultrasonication in acetone, isopropanol and piranha solution ($H_2SO_4$:$H_2O_2$ (30%) 2:1 at 150 °C, 2 hours). This sample was then placed without any further modification in the DBM. For the gas supply (either $N_2$, $NF_3$, or $NH_3$), the GIS was connected to an external gas line, through a connector at the capillary. The gas was then delivered to the capillary through a gas line, where a liquid nitrogen cold trap was used to minimize residual water content in the gas stream delivered to the sample.[27] Figure 1 (b) shows a charge coupled device (CCD) image of the setup during the experiment, where the GIS needle was placed within 500 μm of the sample surface for optimal gas flux at the surface. During the experiments the background chamber pressure increased from 1.3 x $10^{-6}$ mBar to 9.0 x $10^{-5}$ mBar. We note that the pressure increase is measured at a peripheral point relative to the substrate – however, the local pressure increase in the vicinity of the processing area is known to be substantially higher, as typically used for applications in electron/ion beam induced deposition (IBID).[25] We then used a (4.7 ± 0.3) pA, 30 keV $Ar^+$ beam to initiate the recoil implantation of the gas adsorbate. We note that we used $Ar^+$ instead of $Xe^+$, because the momentum transfer is maximized when the atoms are similar in mass, and thus can achieve better implantation conditions.[16] Furthermore, Ar is lighter than Xe, causing less collateral damage to the host crystal. Unlike Xe, Ar is not known to produce any luminescent color center in diamond, hence it should act as a truly inert primary ion in this system. By using the scanning and precise timing control capabilities of the FIB system, a straightforward variation of the irradiation fluence was achieved by altering the number of passes per unit area scanned by the beam. In each pass, the beam is scanned in a serpentine pattern with a Dwell Time of 200 ns, a Point Pitch of 200 nm, and a Defocus of 50 μm, which yields a fluence of ~ 1.7 x $10^{10}$ $cm^{-2}$ for a single pass. Due to the volatile nature of the adsorbates, we also applied a refresh time, which was set to 10 ms between passes in order to facilitate sufficient time for the precursor molecules to replenish around the processing site.[28] Specifically, as the ion beam irradiates the sample, the precursor gas may either be implanted by recoil, auto-desorb or undergo stimulated desorption caused by the ions and secondary electrons emitted from the substrate.[29] Hence, during FIB radiation, adsorbates at the implantation site are depleted and their concentration has to be replenished continuously.[30] We emphasize here that for this version of recoil implantation, no deposition of a thin film in any form is required, which also removes the need for post-implantation sample treatments that are needed to remove such films.

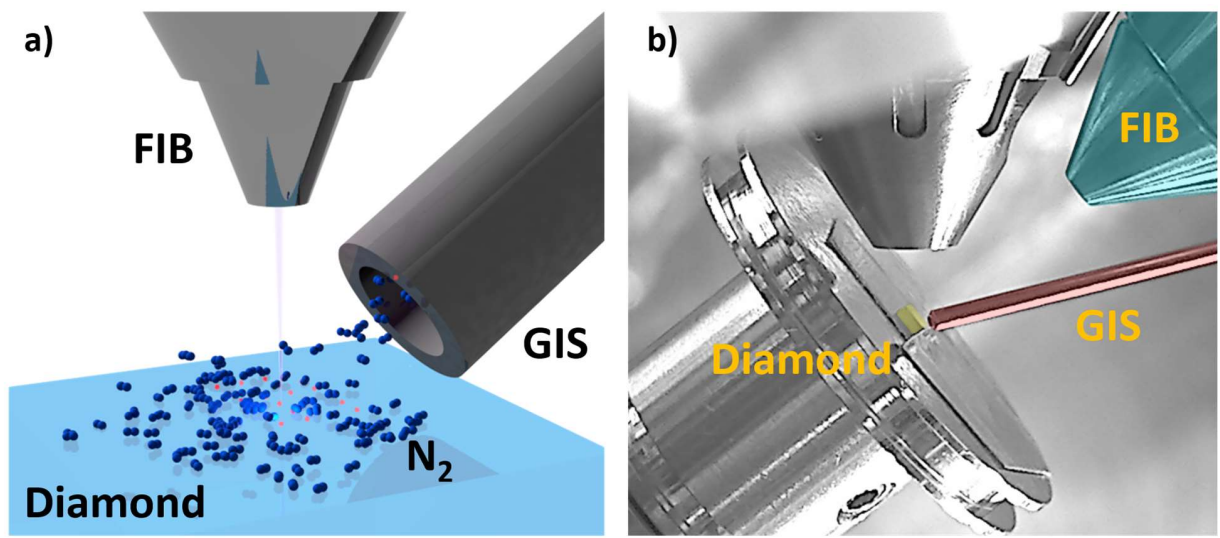

*Figure 1. Experimental setup. a) Schematic illustration of the setup used in the experiments with stylized $N_2$ molecules and a focused ion beam. b) CCD image of the experimental setup used for all implantations with false coloring applied to labelled components.*

After irradiation, the sample was annealed at 850 °C for 2 hours in high vacuum (< 2 x $10^{-6}$ mBar) and cleaned in piranha solution as described above. A lab-built confocal photoluminescence (PL) setup with a 532 nm excitation laser was then used to investigate the irradiated regions at room temperature. The resulting PL map of an area irradiated with the precursor $NF_3$ can be seen in Figure 2 (a). An ion fluence of 7.0 x $10^{13}$ cm$^{-2}$ is required to clearly see the patterned 4 x 4 µm$^2$ squares in the confocal PL map. This is higher than but comparable to our previous work using solid metallic precursors, where a fluence of less than 1 x $10^{12}$ cm$^{-2}$ was needed to observe patterned regions of the same size.[19] The values obtained in this study are higher because of the lower concentration of adsorbates on the surface compared to a solid thin film and the ability for the gaseous precursors to desorb or diffuse unlike the solid metallic precursor. This naturally results in a lower probability of the recoil process, and therefore requires higher total fluence values. Moreover, these values are also significantly higher than using direct nitrogen ion implantation which often require fluences of ~1.0 x $10^{10}$ cm$^{-2}$ as there is an added requirement for momentum transfer between the primary ion and precursor.[4, 5] As is common for all forms of ion implantation, we also observe here, the PL intensity increases up to a maximum and then saturates or declines as ion fluence is increased. The maximum PL intensity is obtained at a fluence of 7.0 x $10^{14}$ cm$^{-2}$ at which point it begins to decrease in the center of the irradiated region. This is related to the increased amount of collisions within the sample, which leads to irreparable damage in the crystal and therefore quenching of the emission.

Figure 2(b) shows individual PL spectra from the square centers, clearly showing the increased intensity of NV$^-$ emission with a pronounced broad phonon sideband (PSB) and a characteristic zero phonon line (ZPL) at ~ 638 nm. Furthermore, in the spectrum a distinct Raman line at 573 nm is observed, corresponding to the $F_{2g}$ mode (1332 cm$^{-1}$) of the sp$^3$ bond, characteristic for the crystal structure of diamond. A further peak is observed at 582 nm, assigned partially to the G band, indicating some degree of damage in the material, as well as the NV$^0$ emission, which is typically present at 575 nm.

We now briefly discuss the non-trivial kinetics arising from the interplay of momentum transfer processes and gas diffusion above and on the surface. First, we note that our choice of a diamond substrate and PL analysis of NV centers prove conclusively that the nitrogen is implanted below the surface and embedded in a diamond crystal – i.e., the ion irradiation did not merely generate a thin film of nitride that can form on some materials as a result of beam-stimulated chemical reactions between adsorbates and a surface. Next, we point out that the appearance of distinct box shapes in Figure 2 (a) with clear edges blurred only by the resolution of the confocal PL imaging system is a direct indicator that the implantation is initiated by momentum transfer to adsorbed gas molecules rather than the less-probable process of momentum transfer to gas-phase molecules above the sample. If the latter was the case, the angular distribution resulting from non-head-on collisions between primary ions and gas molecules would inevitably result in blurred, overlapping implantation sites. Finally, we highlight that the FIB irradiation conditions that we employed do in fact lead to net nitrogen implantation, which is not negated by the net competing effect of sputtering and desorption stimulated by the ions and emitted secondary electrons.

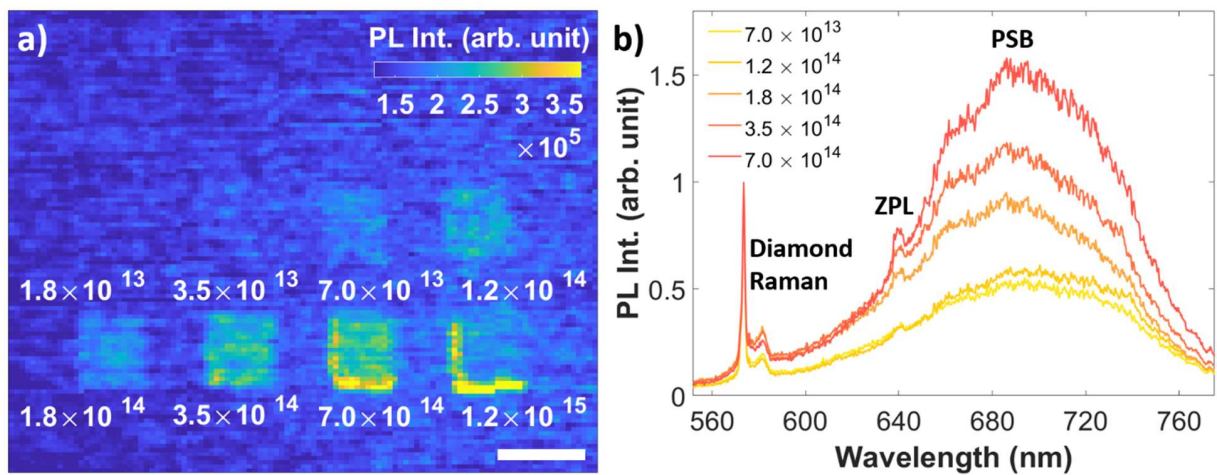

*Figure 2. Photoluminescence map and spectra of implanted arrays of NV$^-$ centers.* *a) A room temperature confocal photoluminescence map of square NV$^-$ arrays implanted using NF$_3$ precursor. The scale bar corresponds to 5 μm. b) Spectra taken from irradiated areas in (a) corresponding to ion fluences in the range of 7.0 x 10$^{13}$ cm$^{-2}$ – 7.0 x 10$^{14}$ cm$^{-2}$. The NV$^-$ zero phonon line (ZPL), phonon side band (PSB) and diamond Raman line are labelled for clarity. The spectra have been normalized to the Raman peak at 573 nm.*

We now turn to a comparison of implantation from different gas species and characterization of the optically detected magnetic resonance (ODMR) signal.[21] First, Figure 3 (a) shows spectra obtained using different nitrogen-based precursors, namely N$_2$, NF$_3$ and NH$_3$, each of which was irradiated using an ion beam fluence of $3.5 \times 10^{14}$ cm$^{-2}$, at a chamber pressure of $9.0 \times 10^{-5}$ mBar. Within the different regions, NF$_3$ showed the brightest PL, followed by N$_2$ and NH$_3$ respectively. These variations in efficacy are likely a result of the complex interplay between residence times of the precursor molecules, ion-adsorbate interaction cross-sections, adsorbate dissociation mechanisms and chemical effects of reactive H and F species. As an example, NF$_3$ and NH$_3$ molecules have longer surface residence times than N$_2$. In turn, it can be expected that NH$_3$ would more likely lead to H-mediated chemical etching of diamond, as it has been shown that hydrogen containing precursors can etch carbon. This may directly explain the order of the PL intensity, with NF$_3$, the highest due to long residence times, while for NH$_3$ the additional etching mechanism leads to slow beam-induced volatilization of the surface.[31, 32] Moreover, due to co-implantation of hydrogen, the NV$^-$ center can be depleted and become optically inactive.[33] Besides the characteristic PL emission we also observed a clear PL contrast when applying a microwave field, a signature of the NV$^-$ optically detectable magnetic resonance (ODMR). Specifically, a microwire was placed in the vicinity (~ 20 µm) of an irradiated square region ($7.0 \times 10^{14}$ cm$^{-2}$ fluence, N$_2$ precursor) and the PL contrast was recorded as the microwave frequency was swept across the range 2.75 – 2.96 GHz. As shown in the inset of Figure 3(a) a PL contrast of ~ 1.5 % was observed at 2.87 GHz, corresponding to the zero-field splitting of the system, separating the electron spin $|0\rangle$ and $|\pm1\rangle$ sublevels of the ground state. We note that the ODMR peak is relatively broad, indicative of strain in the sample. Nevertheless, the ODMR measurement shows the capability of this technique to generate spin defects in diamond.

Finally, to exclude the possibility that the NV$^-$ centers were generated by ion beam processing of native nitrogen impurities in the diamond substrate (N < 5 ppb), or from nitrogen impurities in the Ar$^+$ beam, we compared the implanted regions to areas that were irradiated by the beam in the absence of a precursor gas. As is shown by the blue spectrum in Figure 3(a), the characteristic NV⁻ emission was absent from such regions, irradiated using the same ion fluence, confirming that the observed NV centers were generated through recoil implantation. We did, however, observe bright square arrays in panchromatic PL maps of the areas irradiated in the absence of a precursor gas (Fig. 3(b)). However, these emissions were unstable – focusing the laser spot onto such an irradiated area and taking a further PL map results in a dim spot (Fig. 3(c)). Therefore, we attribute the bright square to carbon deposition via an IBID process stemming from residual hydrocarbon species on the sample surface.[25] The laser dwelling removes the contamination, which leads to PL quenching in that spot. This illustrates that the low ion flux used for the irradiation resulted in a favorable regime between the competing processes of sputtering, implantation and desorption/replenishment.

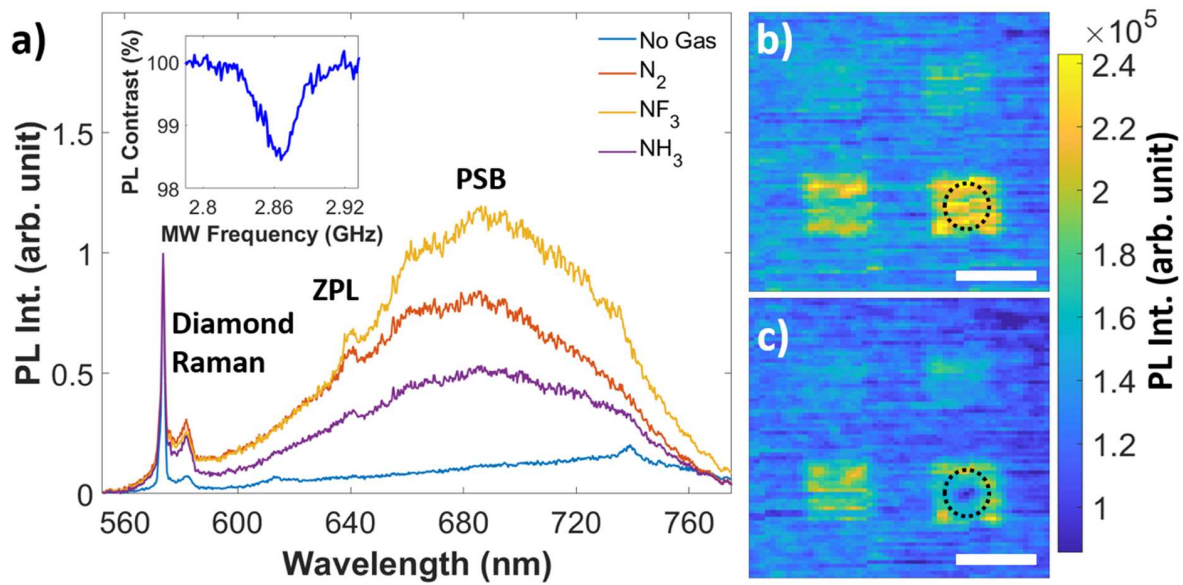

*Figure 3. Comparison of photoluminescence spectra using different nitrogen-based precursors. a) Room temperature photoluminescence spectra of the implanted regions using a fluence of 3.5 x 10$^{14}$ ions cm$^{-2}$. A reference spectrum of a region irradiated in the absence of a precursor gas is*

*shown for comparison. The NV⁻ zero phonon line (ZPL), phonon side band (PSB) and diamond Raman line are labelled for clarity. The spectra are normalized to the diamond Raman line at 573 nm. (Inset) ODMR contrast measurement of the area irradiated with nitrogen gas precursor. b, c) Confocal panchromatic PL maps of regions generated by ion beam irradiation in the absence of a precursor gas obtained before and after the laser spot was dwelled over the circled area for 2 mins. Scale bars correspond to 5 µm.*

In summary, we used a succinct set of experiments – namely, site-selective fabrication of NV⁻ centers in diamond using three precursor gases – to show that:

- Optically-active defects can be fabricated in the substrate lattice below the sample surface.
- The implantation proceeds through momentum transfer to surface-adsorbed gas molecules.
- Our irradiation conditions lead to implantation at a rate that overcomes the competing processes of autodesorption, stimulated desorption and sputtering.
- The ion beam scanning can be used to control both the location and the local dose of the implanted species.
- Both inert and reactive gases can be used as implantation precursors.

**CONCLUSION**

We have extended the range and suitability of the recoil implantation technique and demonstrated its applicability to elements which are available from a gaseous molecular state. By delivering gas-phase molecules to the substrate surface by means of a GIS capillary we achieve a region of locally high adsorbate concentrations, which acts as a source of target atoms that can be implanted through momentum transfer from an energetic ion beam. The principle of this extension was proven by patterning regions of optically-active NV⁻ centers in bulk diamond, by flowing a nitrogen containing gas over the surface during ion beam patterning. After annealing, the characteristic signature of the NV⁻ color center, including the spectrum and ODMR signal with zero-field splitting at 2.87 GHz were observed. Beyond the expansion recoil implantation to a wider range of the periodic table, we have also simplified the material preparation steps by utilizing a gas phase precursor, where no chemical removal of the target film after implantation is required. Such a process is therefore more appealing to materials that are chemically unstable or would be affected by the wet chemical removal process after ion irradiation.

## ACKGNOWLEDGEMENTS

The authors thank the Australian Research Council (DP180100077, DP190101058, LP170100150, CE200100010) for financial support. We would also like to thank Mark Lockrey for his assistance with the Focused Ion Beam microscope.